\documentclass{PoS}
\usepackage[caption=false]{subfig}

\title{A low-luminosity type-1 QSO sample\\- A morphological study of nearby AGN hosts - }

\ShortTitle{Morphological study of nearby AGN hosts}

\author{\speaker{Gerold Busch}$^1$, Jens Zuther$^1$, M\'onica Valencia-S.$^1$, Lydia Moser$^1$, and Andreas Eckart$^{1,2}$\\
        $^1$ I. Physikalisches Institut, Universit\"at zu K\"oln, Z\"ulpicher Str. 77, 50937 K\"oln, Germany\\
        E-mail: \email{busch@ph1.uni-koeln.de}\\
        $^2$ Max-Planck-Insitut f\"ur Radioastronomie, Auf dem H\"ugel 69, 53121 Bonn, Germany}

\abstract{There is growing evidence that every galaxy with a considerable spheroidal component hosts a supermassive black hole (SMBH) at its
center. Strong correlations between the SMBH and the spheroidal component suggest a physical connection through a coevolutionary scenario.

For very massive galaxies a merger-driven scenario is preferred, resulting in elliptical galaxies. In the nearby universe, we find many disk galaxies, showing no signs of recent interaction. Alternative secular evolutionary scenarios for such galaxies involve internal triggers like bars and spiral arms or minor mergers.

We analyze a sample of 99 nearby galaxies ($0.02<z<0.06$) from the Hamburg/ESO survey in order to get insight into structural and dynamical properties of the hosts to trace the origin of the bulge-SMBH correlation. 

In this work, we first collect images of sample members to get an impression of the morphological distribution in the sample. In a second step, we start to analyze sensitive, high resolution near-infrared images of 20 galaxies, performing aperture photometry and bulge-disk decomposition with the BUDDA code. We find an unexpected high fraction of barred galaxies and many other structural peculiarities.}

\FullConference{Workshop on: Nuclei of Seyfert galaxies and QSOs - Central engine \& conditions of star formation\\
		November 6-8, 2012\\
		Max-Planck-Insitut f\"ur Radioastronomie (MPIfR), Bonn, Germany}

\begin{document}

\section{Introduction}
Luminous quasi-stellar objects (QSO), residing in very massive ellipticals, are believed to originate in the violent gravitational interaction of massive spiral galaxies (mergers) \cite{sanders88}. For less massive galaxies, the importance of internal (secular) and external (merger) processes of galaxy evolution is discussed \cite{kormendy04}. Strong correlations between the supermassive black holes (SMBH), believed to reside in the centers of most galaxies \cite{kormendy95}, and the properties of the host galaxy's central spheroidal component \cite{magorrian98,ferrarese00,gebhardt00,marconi03,graham07} suggest a coevolutionary scenario. The role of central activity in the context of galaxy evolution is still unclear.

We observed low-z low-luminosity type-1 QSOs just below the classical Seyfert/QSO demarcation. They are more luminous than other nearby AGN but still close enough to resolve structures of the host galaxy. We hope for insight into structural and dynamical properties to ultimately obtain information on the origin of bulge-SMBH correlation and feedback-mechanisms.

\begin{figure}[b]
\centering
\includegraphics[width=0.6\linewidth]{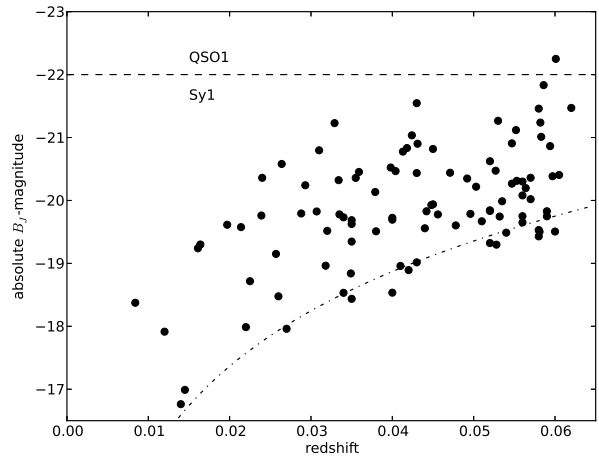}
\caption{Redshift-magnitude diagram of the galaxies in the low-luminosity QSO sample. It is shown that the galaxies lie just below the classical Seyfert-QSO demarcation at $-22^{\textrm{mag}}$. The dashed-dotted line indicates the flux limit of the HE-survey.}
\label{fig:redsh-magn}
\end{figure}

Our sample is a representative subsample of the Hamburg/ESO survey \cite{Wisotzki00} that is a wide angle survey for optically bright QSOs with a flux limit of $B_J\leq17.3$. Our subsample contains only the nearest QSOs: We chose only objects with a redshift z$\leq$0.06. This redshift limit is based on NIR spectroscopic constraints: It ensures the presence of the CO(2-0) rotation vibrational band head absorption line in the K-band. These selection criteria result in 99 nearby low-luminosity type-1 QSOs (see the distribution of absolute magnitudes and redshifts in Fig. \ref{fig:redsh-magn}) \cite{Betram07,fischer06,koenig09,koenig12}.

\section{Initial statistics}
So far, we have high quality near-infrared images, obtained with SofI (NTT), ISAAC, NACO (VLT) and LUCI (LBT), of 27 galaxies. Additionally, we find optical images of another 19 galaxies in the SDSS database, resulting in 46 galaxies, roughly 50\% of our sample.

\begin{figure}
\centering
\includegraphics[width=0.6\linewidth]{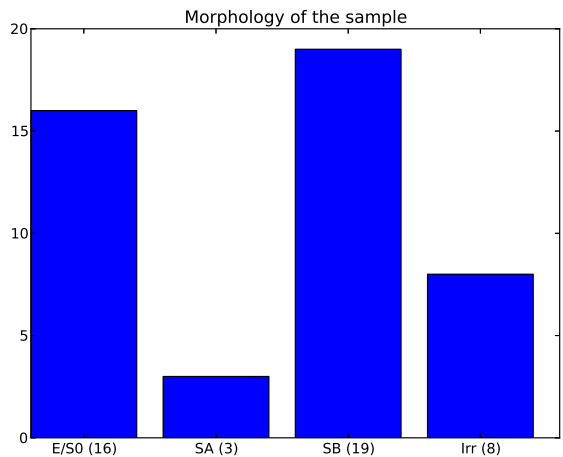}
\caption{Initial statistics of 46 of the 99 galaxies from the sample. We distinguish between circular/elliptical galaxies (E/S0), spiral galaxies with (SB) and without (SA) bar and irregular galaxies. The classification is based on visual inspection and decompositions if available. Many of the galaxies classified as SA, SB or E/S0 show peculiarities. However, we only classified them as Irr if no clear classification of the extended underlying structure was possible.}
\label{fig:statistics}
\end{figure}

An initial statistic shows that a small but significant fraction ($\approx$10\%) of them is interacting, e.g. as major merger. The non-interacting galaxies are mostly spirals. The classification is presented in Table \ref{tab:othergals}. Fig. \ref{fig:statistics} shows the distribution of host galaxy morphologies. It is interesting that in the sample presented here the spiral galaxies are almost all hosting a bar (19 out of 22). The fraction of barred galaxies is higher than expected (e.g. \cite{kormendy04}: bar detection rate of one third at optical wavelengths and two thirds in the infrared). This finding requires more detailed investigation of the role of bars in the context of the secular evolution of nearby AGN host galaxies.

\begin{table}
\centering
\caption{Overview of morphological classification. [a] galaxies analyzed in our NIR study (Busch et al., 2013, in prep.), [b] analyzed in a former study \cite{fischer06}, abbreviations: E=bulge dominated galaxy, SA=spiral galaxy without bar, SB=spiral galaxy with bar, Irr=Irregular}
\label{tab:othergals}
\begin{tabular}{cccc} \hline \hline
Name & Instrument/ & Classi- & Notes \\ 
 & Survey & fication & \\ \hline
05HE0036-5133 & SofI & E & [a] \\
06HE0038-0758 & SDSS & SB & --- \\
07HE0040-1105 & SDSS & E & --- \\
08HE0045-2145 & SofI & SB & [a] \\
11HE0103-5842 & SofI & Irr & [a] \\
13HE0108-4743 & NACO & SB & --- \\
15HE0114-0015 & SDSS & SB & --- \\
16HE0119-0118 & SofI & SB & [a] \\
19HE0126-0753 & SDSS & SB & --- \\
21HE0150-0344 & SDSS & Irr & interacting \\
22HE0203-0031 & SDSS & Irr & disturbed \\
23HE0212-0059 & SDSS & SA & --- \\
24HE0224-2834 & SofI & Irr & [a] \\
25HE0227-0913 & SDSS & E & --- \\
26HE0232-0900 & SDSS & Irr & disturbed \\
29HE0253-1641 & SofI & SB & [a] \\
32HE0330-1404 & SDSS & SB & --- \\
37HE0345+0056 & SDSS & E & --- \\
50HE0444-0513 & SDSS & SB & --- \\
51HE0447-0404 & SDSS & Irr & disturbed \\
55HE0853-0126 & SDSS/ISAAC & SB & [b] \\
56HE0853+0102 & SDSS & Irr & --- \\
57HE0934+0119 & SDSS & SB & --- \\
58HE0949-0122 & SDSS & E & --- \\
60HE1013-1947 & ISAAC & SB & [b] \\
61HE1017-0305 & SDSS/ISAAC & SB & nearby companion, [b] \\
62HE1029-1831 & ISAAC & SB & [b] \\
69HE1248-1356 & LUCI & SA & [a],[b] \\
70HE1256-1805 & LUCI & E & [a] \\
71HE1310-1051 & LUCI & E & [a], disk? \\
73HE1328-2508 & ISAAC & Irr & interacting,[b] \\
74HE1330-1013 & LUCI & SB & [a] \\
75HE1338-1423 & LUCI & SB & [a],[b] \\
77HE1348-1758 & LUCI & E & [a] \\
79HE1417-0909 & LUCI & E & [a] \\
\hline \\
\end{tabular}
\end{table}

\begin{table}
\centering
\ContinuedFloat
\caption{cont.}
\begin{tabular}{cccc} \hline \hline
Name & Instrument/ & Classi- & Notes \\ 
 & Survey & fication & \\ \hline
80HE2112-5926 & SofI & E & [a] \\
81HE2128-0221 & SofI & E & [a] \\
82HE2129-3356 & SofI & E & [a] \\
83HE2204-3249 & SofI & E & [a] \\
84HE2211-3903 & SofI & SB & [a],[b] \\
85HE2221-0221 & SofI & E & [a] \\
86HE2222-0026 & SDSS & E & --- \\
88HE2233+0124 & SDSS & SB & --- \\
89HE2236-3621 & SofI & E & [a] \\
93HE2302-0857 & SDSS & SA & --- \\
99HE2354-3044 & NACO & SB & interacting \\
\hline \\
\end{tabular}
\end{table}

\section{Decomposition}
A first qualitative structural overview is obtained through visual inspection. In a next step we use two dimensional decomposition as a more quantitative approach.
We used the BUlge Disk Decomposition Analysis (BUDDA) Code \cite{budda} to separate disk, bulge, bar and nuclear component (AGN). We determine the fractions of the individual components and hope to reveal otherwise hidden substructures. As useful for most galaxies, BUDDA fits the disk with an exponential law. Bulge and bar are fitted with a S\'ersic-function using a S\'ersic-index around 4 (bulge) or around 0.7 (bar) \cite{gadotti08}. The AGN was modelled by an unresolved point source, convolved with a Moffat distribution, describing the PSF. The other components are smoothed with the same function in order to account for the seeing.

\begin{figure}
\centering
\subfloat[]{
\includegraphics[width=0.48\linewidth]{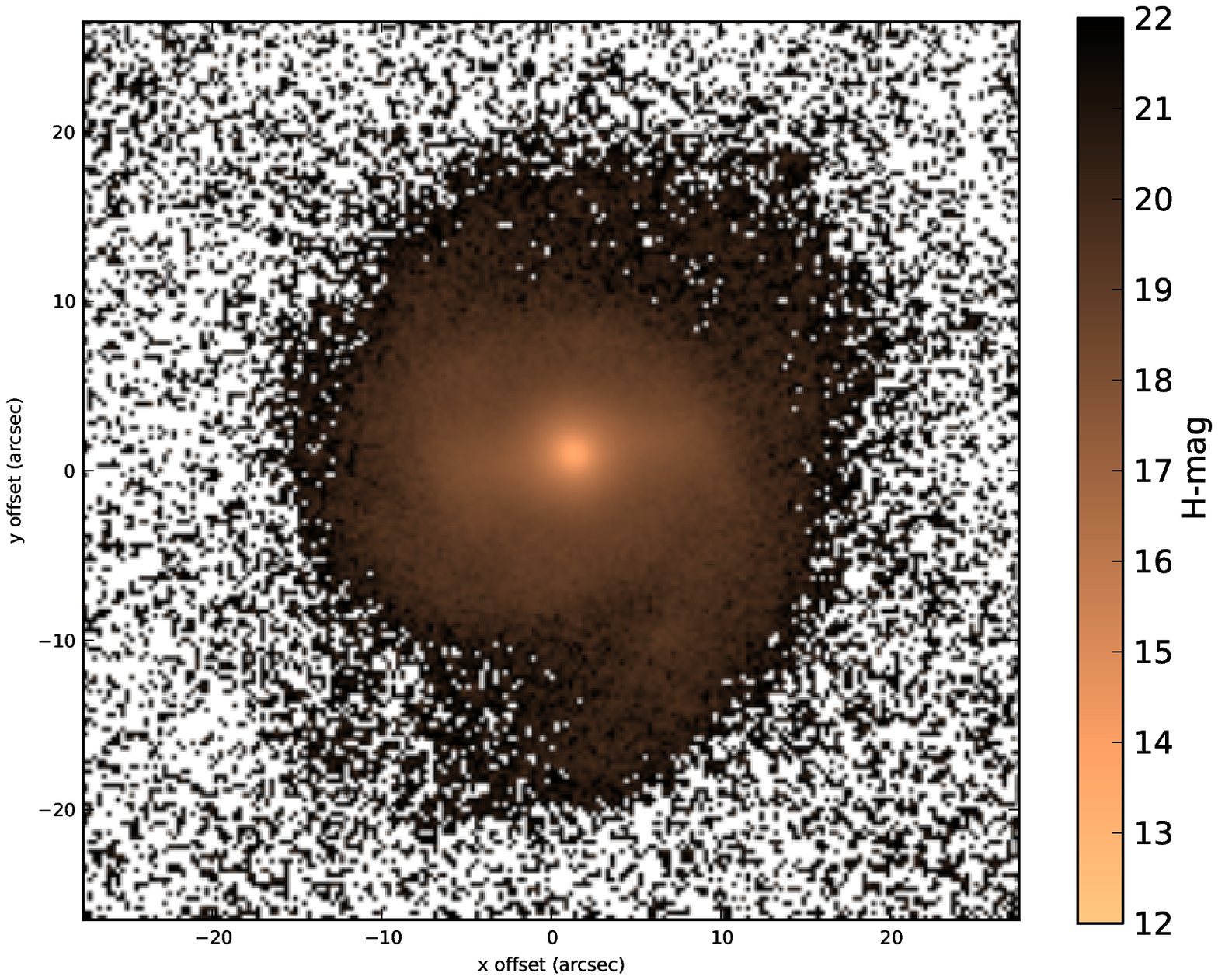}}
\subfloat[]{
\includegraphics[width=0.48\linewidth]{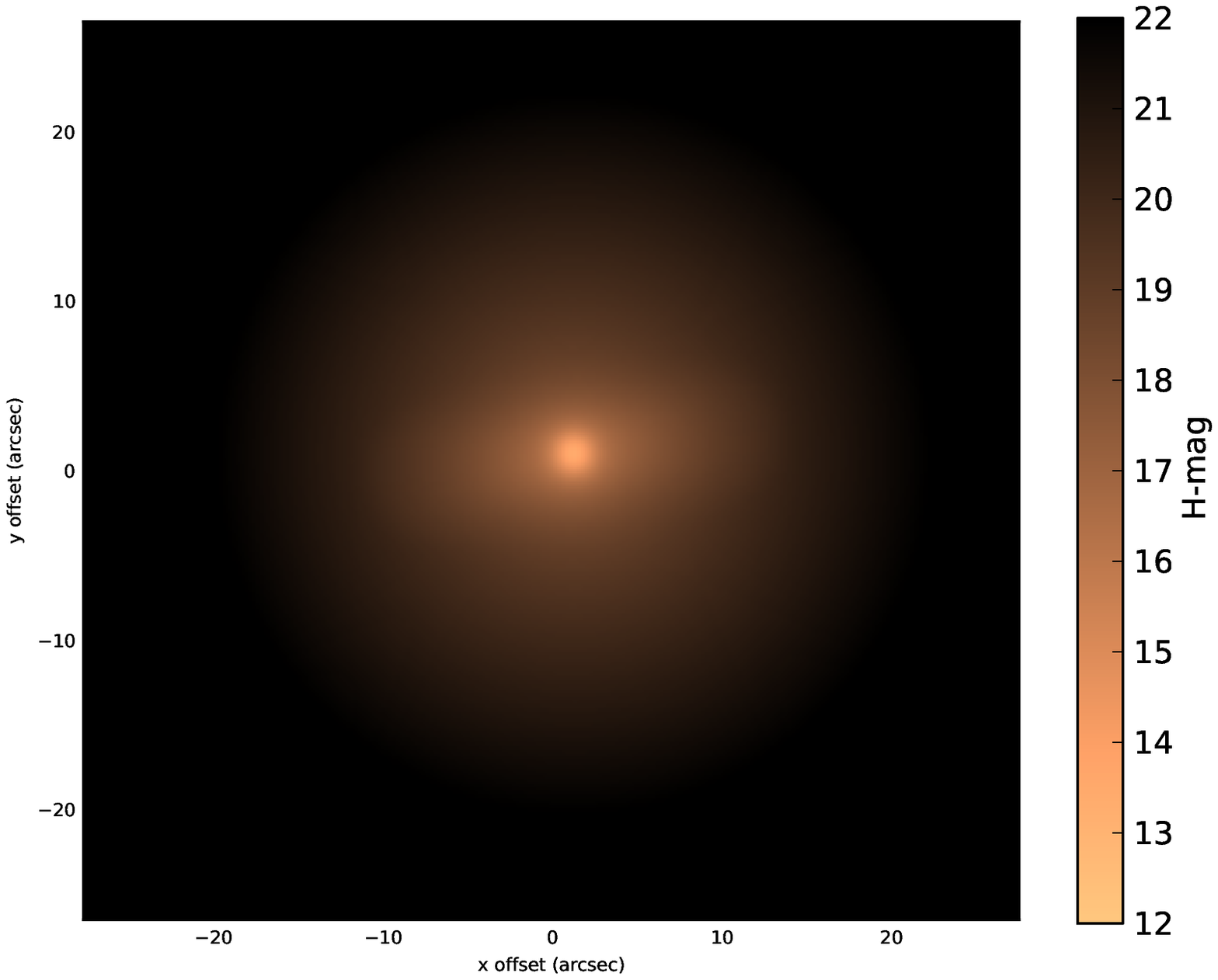}}\\
\subfloat[]{
\includegraphics[width=0.48\linewidth]{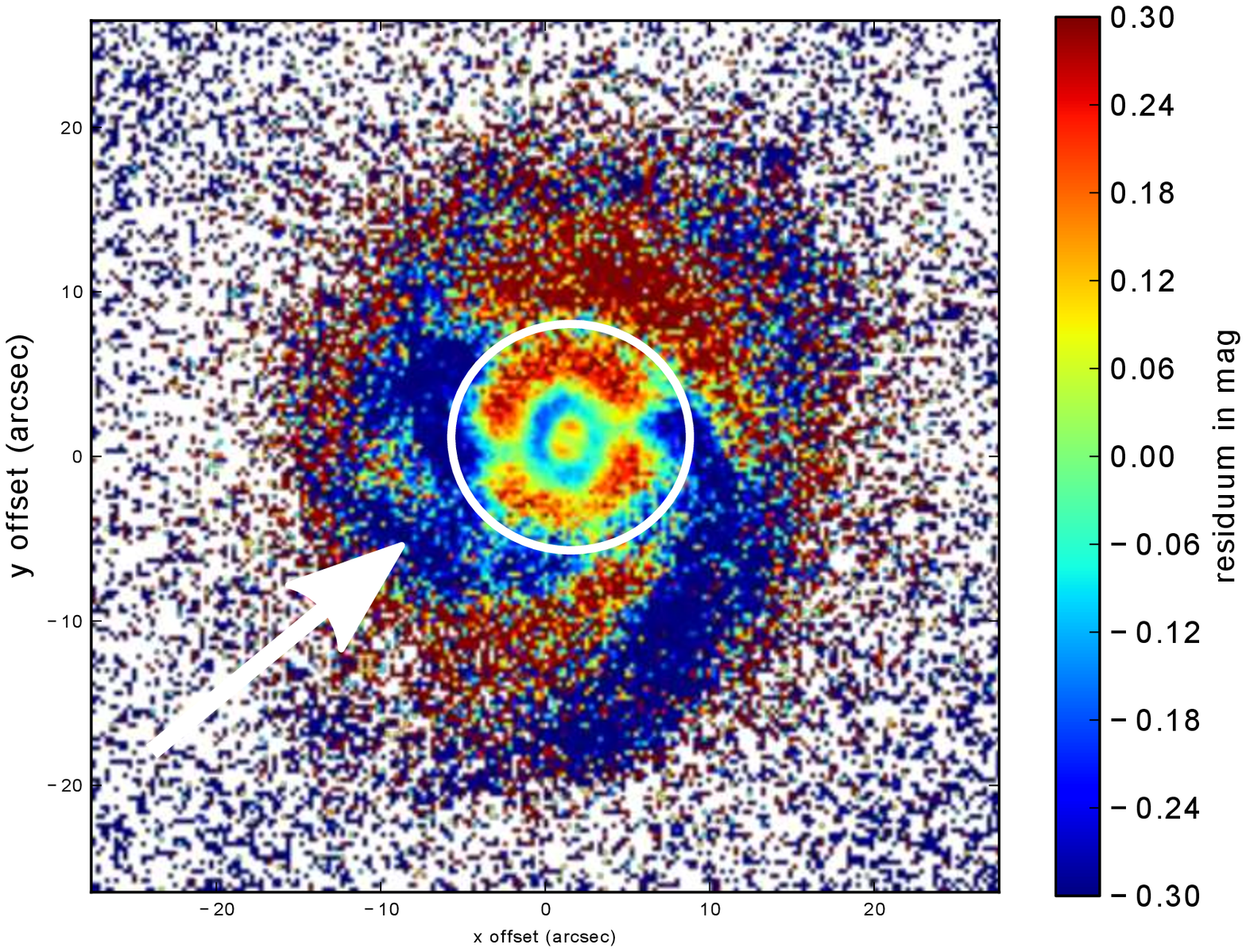}}
\subfloat[]{
\includegraphics[width=0.48\linewidth]{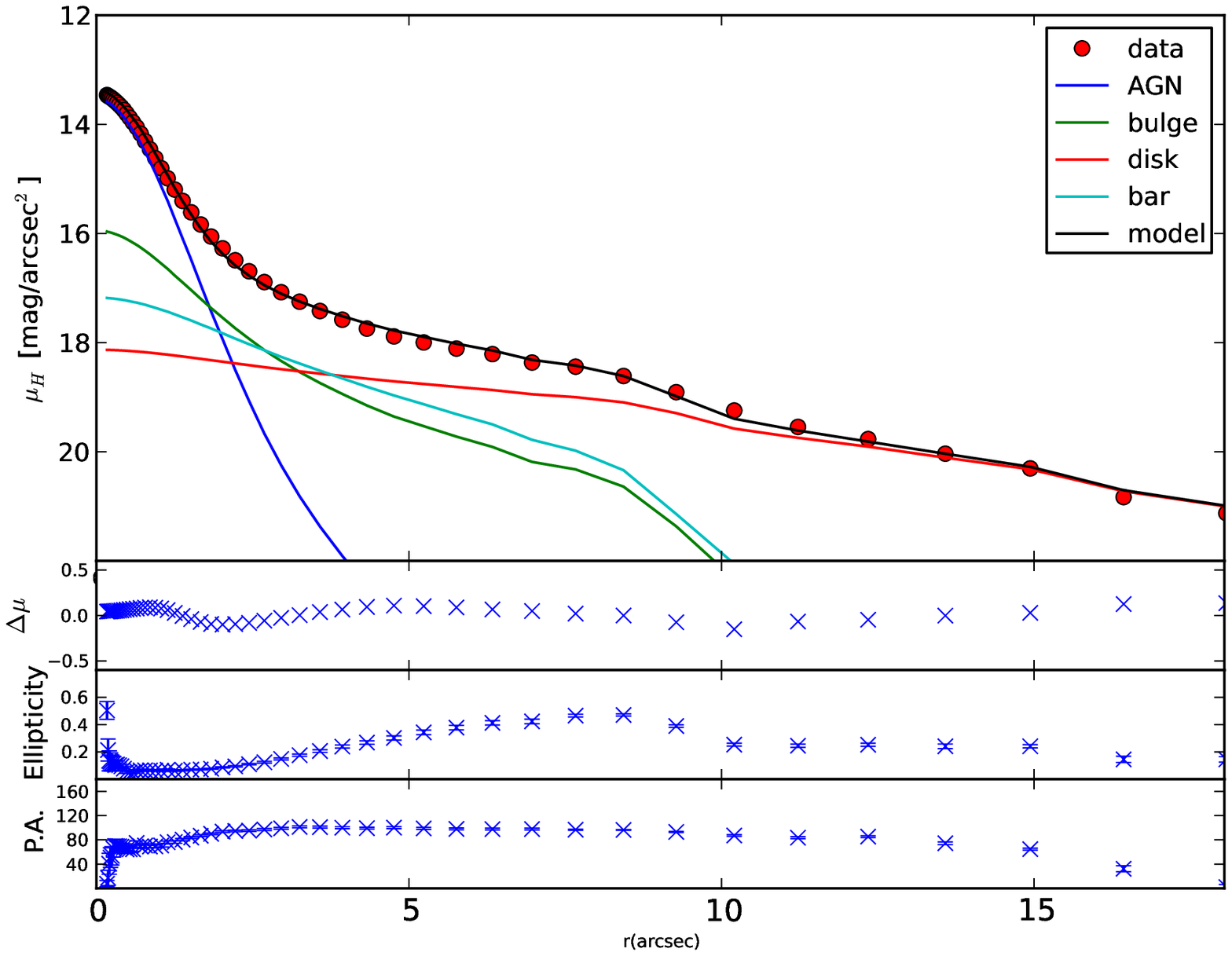}}
\caption{Decomposition of galaxy HE2211-3903: (a) H-band image, (b) model image, (c) residuum, obtained by dividing original/model, (d) radialprofiles of H-band manitudes of data, single components and model, difference between data and model, ellipticity and position angle (averaged over ellipses). Substructures mentioned in the text are marked in (c).}
\label{fig:decomposition}
\end{figure}

Fig. \ref{fig:decomposition} shows an example. The residuum shows an inner ring and a third spiral arm as substructures that were otherwise hidden (indicated by a circle and an arrow in Fig. \ref{fig:decomposition} (c) ). They can be approved by line maps \cite{scharw12}. We use the derived parameters for detailed structural analysis of the observed galaxies and to gain information about the host galaxy's properties as e.g. black hole mass estimates \cite{marconi03}.

The detailed results of the NIR study will be presented in Busch et al. (2013, in prep.).

\vspace{7mm}
\begin{small}
\paragraph{Acknowledgements}
We thank Dimitri Gadotti for making BUDDA publicly available and helpful support. Further, we thank Lutz Wisotzki for valuable comments and advice concerning the Hamburg/ESO survey and our QSO sample.  \\
GB is member of the Bonn-Cologne Graduate School of Physics and Astronomy and acknowledges support from Konrad-Adenauer-Stiftung e.V. JZ acknowledges support by the German Academic Exchange Service (DAAD) under project number 50753527 and by BMBF 05A08PKA. This work has in part been supported by the German Deutsche Forschungsgemeinschaft, DFG, via grant SFB 956.

\end{small}
\end{document}